\documentclass[%
reprint,
superscriptaddress,
nofootinbib,
 amsmath,amssymb,
 aps,
]{revtex4-2}

\usepackage{graphics,epsfig,psfrag}
\usepackage{color}
\usepackage{xcolor}
\usepackage{graphicx}
\usepackage{dcolumn}
\usepackage{bm}
\usepackage{hyperref}

\usepackage{mathrsfs}

\hypersetup{
           breaklinks=true,   
           colorlinks=false,   
           pdfusetitle=true,  
        }

\begin{document}

\preprint{APS/123-QED}

\title{Constraining spherically symmetric metrics by the gap between photon rings}%

\author{Fabio Aratore}
\email{faratore@unisa.it}
\affiliation
{Dipartimento di Fisica “E.R. Caianiello”, Università degli studi di Salerno, Via Giovanni Paolo II 132, I-84084 Fisciano SA, Italy}
\affiliation{Istituto Nazionale di Fisica Nucleare (INFN), Sezione di Napoli - Gruppo collegato di Salerno, Via Cintia, 80126 Napoli NA, Italy}

\author{Oleg Yu. Tsupko}
\email{tsupkooleg@gmail.com}
\affiliation{ZARM, University of Bremen, 28359 Bremen, Germany}

\author{Volker Perlick}
\email{perlick@zarm.uni-bremen.de}
\affiliation{ZARM, University of Bremen, 28359 Bremen, Germany}

\date{\today}

\begin{abstract}

Gravitational lensing of luminous matter that surrounds a black hole or some other sufficiently compact object produces an infinite sequence of images. Besides the direct (or primary) image, it comprises demagnified and deformed replicas of the original known as photon rings which are progressively nearing the boundary of the so-called shadow. In the present paper, we present analytical approximation formulas for higher-order photon rings for an asymptotically flat, static, spherically symmetric spacetime that admits a photon sphere. We consider an emission ring in the equatorial plane and an observer at arbitrary inclination far away from the center. Fixing the emission radius and leveraging the strong deflection limit, which provides an analytical logarithmic approximation for the deflection angle, we find the deformed shape of higher-order photon rings in the form of a polar equation on the observer’s screen. It has been suggested by other authors to use the relative size of photon rings for characterizing the underlying spacetime. In particular, the relative separation between two neighboring photon rings, which we call “gap parameter”, was considered. We obtain an analytical expression for the gap parameter of higher-order photon rings for metrics of the considered class that may depend on multiple parameters.  The advantage of using this quantity is in the fact that, to within the assumed approximations, it is independent of the mass of the central object (or of some other characteristic parameter if the mass is zero) and of the distance of the observer. Measurements of the gap parameter, which may become possible in the near future, will restrict the spacetime models that are in agreement with the observations. Even without knowledge of the emission radius, it will conclusively rule out some metrics. We exemplify our calculations of the gap parameter with the Schwarzschild, Reissner-Nordstr{\" o}m, Janis-Newman-Winicour and Ellis wormhole metrics. In the second and third cases some coefficients have to be calculated numerically.
\end{abstract}

\maketitle


\section{Introduction}
\label{sec:introduction}

Black holes and other sufficiently compact objects can cause arbitrarily large deflections of light rays. In particular, photons can loop around a black hole arbitrarily many times moving in the vicinity of a \textit{photon sphere}. 

This leads to the formation of an infinite sequence of images of each light source. These images can be labeled by a natural number 0, 1, 2, 3, etc., called its \emph{order}, see more details in the next section. The image of order 0 is called the \emph{primary} (or \emph{direct}) image, the image of order 1 is called the \emph{secondary} image and all other images are called \emph{higher-order} images. (For an illustration of images of a thin accretion disk, see, e.g., Fig.~1 in Ref.{\cite{BK-Tsupko-2022}}.) In the case of perfect alignment the observer sees an infinite sequence of \emph{Einstein rings} \cite{MTW-1973, Ohanian1987, Virbhadra-2000, bozza2001g, bozza2002gravitational, BK-Tsupko-2008}. (Einstein rings are sometimes called \emph{Chwolson rings}, referring to a pioneering paper by Chwolson \cite{Chwolson-1924}.) The existence of a photon sphere and the formation of higher-order images are closely related to the fact that a black hole or another sufficiently compact object casts a \emph{shadow} \cite{Falcke-2000, Bronzwaer-Falcke-2021}: it displays a dark disk on the sky of an observer whose boundary corresponds to past-oriented light rays from the observer position that spiral towards the photon sphere, for reviews see  \cite{perlick2022calculating, Cunha-Herdeiro-2018}. With increasing order the images of a light source become fainter and fainter and approach the boundary of the shadow.

The existence of a photon sphere and the resulting observational phenomena have firstly been discussed for the Schwarzschild metric: In this case the existence of a photon sphere, at $r=3m$ in standard notation, is known since the early days of general relativity, see Hilbert \cite{Hilbert1917}. Much later, Darwin \cite{darwin1959gravity} observed that this leads to the formation of an infinite sequence of images and that the deflection angle diverges logarithmically for light rays that approach the photon sphere. Zeldovich and Novikov \cite{ZeldovichNovikov1966} and Synge \cite{Synge1966} independently calculated what we now call the shadow, for an observer anywhere outside of the horizon;  for details, see Section 2 of review \cite{perlick2022calculating}. In a pioneering paper, Luminet \cite{Luminet1979} calculated and visualized not only what we now call the shadow but also the images of luminous rings, also see Bao \textit{et al}. \cite{Bao-Hadrava-1994a, Bao-Hadrava-1994b}. The images of such extended light sources form rings around the shadow which, for order $n \ge 1$, have recently been termed \emph{photon rings}  (which should not be confused with ``Einstein rings''). Ohanian \cite{Ohanian1987} observed that with increasing order the intensity of images decreases exponentially. More recently, Virbhadra and Ellis \cite{Virbhadra-2000} investigated images of arbitrary order in terms of a lens equation. They introduced the term ``relativistic images'' for the higher-order images which afterwards was used also by some other authors.

For an unspecified spherically symmetric and static spacetime, we refer to Atkinson \cite{Atkinson-1965} who was the first to give the necessary and sufficient condition for the existence of a photon sphere and to Perlick \cite{Perlick-2004-exact-equation} who discussed the formation of images of arbitrary order in terms of an exact lens map. The properties of higher-order images have been discussed for many specific spherically symmetric and static spacetimes, see e.g. Eiroa \textit{et al}. \cite{eiroa2002reissner} for the Reissner-Nordstr{\"o}m spacetime and Perlick \cite{Perlick-2004-review} for a review that includes other examples.\\

The long-standing concept of higher-order images has become of increased importance after the shadow of black holes has actually been observed recently
\cite{akiyama2019first1, akiyama2019first2, akiyama2019first3, akiyama2019first4, akiyama2019first5, akiyama2019first6, Kocherlakota-2021, EHT-SgrA-2022-01, EHT-SgrA-2022-02, EHT-SgrA-2022-03, EHT-SgrA-2022-04, EHT-SgrA-2022-05, EHT-SgrA-2022-06}. 
In particular, photon rings, i.e., higher-order images of extended light sources, such as accretion disks, have been studied quite comprehensively, both analytically and numerically, in Refs. \cite{gralla2019black, johnson2020universal, gralla2020observable, gralla2020shape, Gan-Wang-2021, Hadar-2021-photon-rings, broderick2022photon, Paugnat-2022-photon-rings-shape, Andrianov-2022, Papoutsis-2023-photon-rings, Staelens-2023-photon-rings, Deich-Yunes-2023-photon-rings} and others mentioned later in the present paper. Resolving photon rings around the shadow of a black hole is one of the ambitious goals of planned future observations \cite{johnson2020universal, pesce2021toward, Johnson-2023-Galaxies, Ayzenberg-arxiv-2023}. We note that sometimes the whole sequence of images $n \ge 1$ together is referred to as a ``photon ring"; see the detailed discussion of different terminologies on page 7 of review \cite{perlick2022calculating}.

Although the research on photon rings is rather focused on rotating black holes, particular attention has also been paid to spherically symmetric metrics to which we will restrict in the current paper. Photon rings of an infinitely thin emission ring around different spherically symmetric black holes were recently investigated numerically by Wielgus \cite{wielgus2021photon}. Different spherically symmetric metrics have been considered in a series of works by Guerrero \textit{et al}. \cite{Guerrero-2021, Guerrero-2022a-photon-rings, Guerrero-2022b-photon-rings}, including the interesting case of spacetimes with two photon spheres \cite{Guerrero-2022a-photon-rings}. A comprehensive study of photon rings in different spherically symmetric metrics and a wide range of emission profiles was presented by Kocherlakota \textit{et al}. \cite{Kocherlakota-2023-photon-rings,Kocherlakota-2024-PRD-photon-rings, Kocherlakota-2024-arxiv-photon-rings}. Photon rings in the case of alternative spherically symmetric geometries with thin accretion disks are modeled in Dias da Silva \textit{et al}. \cite{da-Silva-2023-photon-rings}. Spherically symmetric spacetimes are considered by Broderick \textit{et al}. \cite{Broderick-Salehi-2023-photon-rings}, including an analytical analysis of photon rings.

An analytical approximation formula for calculating the deflection angle and higher-order images, called the \emph{strong deflection limit}, was brought forward by Bozza \textit{et al}. \cite{bozza2001g} for the Schwarzschild spacetime and then by Bozza \cite{bozza2002gravitational} for a general spherically symmetric and static spacetime that is asymptotically flat and admits a photon sphere. In these works it was assumed that the light source and the observer are both far away from the center. This formalism gives a simple logarithmic expression for the gravitational deflection angle that is a valid approximation for rays that make at least one full revolution around the center. In addition to the Schwarzschild metric, the strong deflection limit was calculated for several other spherically symmetric and static metrics, see, e.g., \cite{bozza2002gravitational, eiroa2002reissner, Eiroa-2004-retrolensing, Eiroa-2005, Gyulchev-2007, BK-Tsupko-2008, Tsupko-2014, Tsukamoto-2016, Tsukamoto-2021}. Also, the formalism has been extended to the case of  light rays that are influenced by a plasma by Bisnovatyi-Kogan and Tsupko \cite{Tsupko-BK-2013, BK-Tsupko-2017-Universe} and to light rays in the Kerr spacetime by Bozza \textit{et al}. \cite{bozza2001g, bozza2002gravitational, Bozza-2003-kerr, Bozza-2005-kerr, Bozza-2006-kerr, Bozza-Sereno-2006}.

Another generalization is crucial for us: Bozza and Scarpetta \cite{bozza2007strong} obtained formulas for the strong deflection limit in a spherically symmetric and static spacetime where the source (and also the observer) may be at arbitrary position; so this modified formalism can be applied to the situation in which the source position is close to the central object, see also \cite{Bozza2010, Aldi-Bozza-2017, aratore2021decoding}. Using this version of the strong deflection limit, the properties of higher-order photon rings around a Schwarzschild black hole have recently been analytically investigated in Refs \cite{BK-Tsupko-2022, tsupko2022shape}. In particular, an analytic formula for the shape of a high-order ring for the case when the source is a luminous equatorial ring around a Schwarzschild black hole was derived by Tsupko \cite{tsupko2022shape}.\\

\begin{figure*}
\begin{center}
\includegraphics[width=0.90\textwidth]{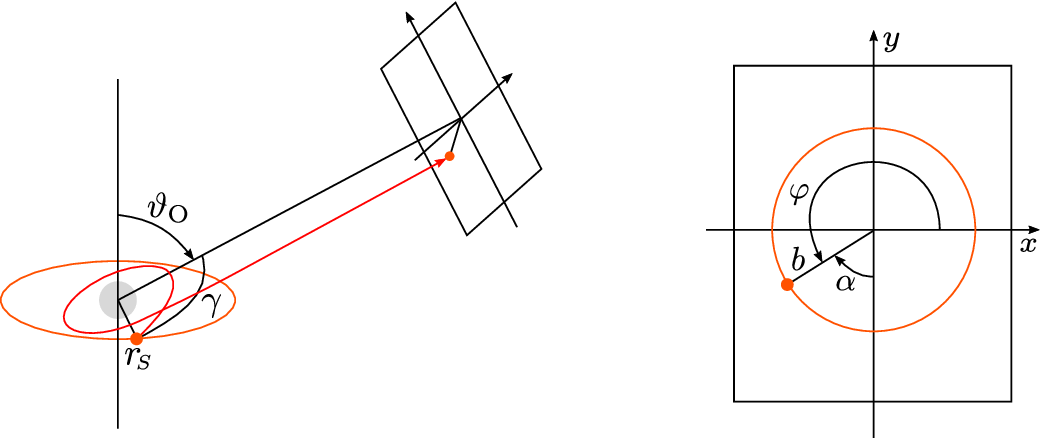}
\end{center}
\caption{
Location of the emission ring and the observer. Left panel: the radiating ring of radius $r_S$ is in the equatorial plane of a spherically symmetric central object. The observer is positioned at a viewing angle $\vartheta_\textrm{O}$. One of the rays forming the tertiary image ($n=2$) is shown. The angle $\gamma$ is measured in the plane of the light ray. Right panel: the picture shows the observer's screen, with polar coordinates $b$ and $\varphi$. The angle $\alpha$ is used in the work of Luminet \cite{Luminet1979}, see Fig.~3 there. In the paper of Tsupko \cite{tsupko2022shape} and in the present paper we transform it to the angle $\varphi$, and the shape of the image is given in polar form, i.e., by the function $b(\varphi)$.
This figure is taken from our previous paper \cite{tsupko2022shape}, see Fig.~2 there.}
\label{fig:geometry}
\end{figure*}

In this paper, we employ the strong deflection limit in the version of Bozza and Scarpetta \cite{bozza2007strong} for computing the shape of higher-order photon rings for an arbitrary asymptotically flat static spherically symmetric metric. As a source, we use the model of a thin ring of given radius, while an observer located at a great distance from the center looks at the emission ring at an arbitrarily given angle (Fig.~\ref{fig:geometry}). The outcome is an analytic expression for the shape of the higher-order image as an explicit formula in polar coordinates on the observer's screen.

Subsequently, we determine, based on the previous results, a characteristic quantity which we call the ``gap parameter" $\Delta _n$. It is defined as the difference between rings of order $n$ and $n+1$ divided by the size of the ring of order $n$. The idea of using the relative separation between photon rings as a probe of the spacetime metric was presented by Eichhorn \textit{et al}. \cite{Eichhorn-2023-photon-rings}, with application to secondary and tertiary images. We also refer to earlier papers of Broderick \textit{et al}. \cite{Broderick-2022-spin} who suggested to use the relative size of photon rings for determining the spin of a Kerr black hole and works of Wielgus \cite{wielgus2021photon} and Ayzenberg \cite{ayzenberg2022testing} where the ratio of radii is analyzed.

In this context, we propose the utilization of the gap parameter for a metric that may depend on multiple parameters. We choose one of them, typically the mass (if it is nonzero), as a characteristic parameter. An analytical expression for the gap parameter for higher-order photon rings is presented. This expression contains coefficients that are given in terms of integrals. While for certain metrics these integrals can be solved in terms of elementary functions, for others a numerical evaluation is needed. The gap parameter is shown to be independent of the characteristic parameter and of the observer's distance. Assuming a luminous disk with given inner and outer radius results in a certain range for the gap parameter $\Delta _n$. Calculating this range for the lowest order, $n=2$, to which our approximations apply can allow one to immediately rule out some metrics, even without knowledge of the inner and outer radii of the emission disk.
Assuming that the emission radius is not known but lies in a given reasonable interval (e.g., from the innermost stable circular orbit (ISCO) to infinity) results in a certain range for the gap parameter $\Delta _n$ (which depends on the emission radius). Calculating this range for the lowest order, $n=2$, to which our approximations apply can allow one to immediately rule out some metrics, even without knowledge of the emission radius or, in case of extended emission, without details of its distribution.

The article is organized as follows: in Sec.~\ref{sec:shape} we recall the basic quantities of lensing in the strong deflection limit and calculate the shape of photon rings in a spherically symmetric and static spacetime. In Sec.~\ref{sec:examples} we apply the general formulas of the previous section to the Schwarzschild, Reissner-Nordstr\"{o}m, Janis-Newman-Winicour and Ellis wormhole metrics. In Sec.~\ref{sec:gapparameter} we introduce the above-mentioned gap parameter and discuss its properties. In Sec.~\ref{sec:gapparameter-analytical} we present an analytical expression for the gap parameter for higher-order images, and we specify it for the four examples from Sec.~\ref{sec:examples}. We draw our conclusions in Sec.~\ref{sec:conclusions}. In an Appendix we compare our notation in the form of a table with the notation used by some other authors.

\section{Shape of higher-order images of a thin equatorial emission ring}
\label{sec:shape}

We consider a static and spherically symmetric spacetime, with a line element of the form
\begin{equation}
ds^2 = - A(r) \, c^2 dt^2 + B(r) \, dr^2 + D(r)  \left( d \vartheta ^2 + \mathrm{sin} ^2 \vartheta \, d \phi ^2\right) \, .
 \label{eq:metric}
\end{equation}
We assume that the metric coefficients $A(r)$, $B(r)$ and $D(r)$ satisfy the conditions of asymptotic flatness,
\begin{equation}
    A(r) \to 1 \, , \quad B(r) \to 1 \, , \quad \dfrac{D(r)}{r^2} \to 1 
\label{eq:asy}
\end{equation}
for $r \to \infty$. This implies that these coefficients are strictly positive for sufficiently large $r$. We denote the maximal interval on which this is true as $\, ] \, r_m  , \infty \, [ \,$.
Here and in the following we use inverted square brackets for an open interval.

We also assume the existence of at least one \emph{photon sphere} in this domain, i.e. a radius value such that a lightlike geodesic stays on the sphere with this radius if it starts tangentially to it.

The equation that determines the radius of a photon sphere, in an arbitrary spherically symmetric and static metric, was first given by Atkinson \cite{Atkinson-1965}, see also \cite{virbhadra2002gravitational, claudel2001geometry, bozza2002gravitational, perlick2022calculating}. In the notation of Eq.~\eqref{eq:metric}, it reads
\begin{equation}
     V'(r_{\mathrm{ph}} ) = 0 \, , \quad \text{where} \quad
    V(r) = - \dfrac{D(r)}{A(r)}  \, .
\label{eq:Atkinson}
\end{equation} 
We allow for the possibility that there are several photon spheres and we denote the radius coordinate of the outermost one by $r_{\mathrm{ph}}$. A photon sphere may be stable or unstable with respect to radial perturbations. An unstable photon sphere lies at a local maximum and a stable photon sphere lies at a local minimum of the potential $V(r)$. The assumption of asymptotic flatness \eqref{eq:asy} implies that $V(r) \to - \infty$ for $r \to \infty$; hence, the outermost extremum of $V(r)$ must be a local maximum, at least from the outer side, so the photon sphere at $r_{\mathrm{ph}}$ is necessarily unstable with respect to outward radial perturbations. For our purpose, we have to restrict to spacetimes where the outermost extremum of $V(r)$ is even an \emph{absolute} maximum of the potential $V(r)$, meaning that the potential is smaller than $V( r_{\mathrm{ph}})$ for all other points in the interval $\, ] \, r_m  , \infty \, [ \,$. Examples of spacetimes which satisfy our assumptions include black holes, where the interval $\, ] \, r_m  , \infty \, [ \,$ extends from the black-hole horizon to infinity, and wormholes, where this interval extends from one asymptotic end to the other. 

Our conditions imply that light rays can asymptotically spiral towards the photon sphere at $r_{\mathrm{ph}}$ from the outer side. This leads to the formation of an infinite sequence of images of each light source. These images can be labeled by a natural number 0, 1, 2, 3, etc., called its \emph{order}, which counts how often the orbit meets the axis, that is the straight line through the observer position and the origin of the coordinate system, not counting the starting point. (This definition is unambiguous and convenient, but it is restricted to spherically symmetric and static spacetimes. In the case of the Kerr spacetime and an observer off the equatorial plane, the images are sometimes labeled by counting how often the corresponding light ray crosses the equatorial plane.)

Because of the symmetry, every light ray lies in a coordinate plane that we parametrize by polar coordinates $(r , \tilde{\phi } )$ where the coordinate $r$ here is the same as in Eq.~\eqref{eq:metric}. It is a standard exercise to calculate the total range of the coordinate $\tilde{\phi}$ that is swept out by a light ray that starts at a light source at radius coordinate $r_S$, goes through a minimum radius value $R$ and then reaches an observer at radius coordinate $r_O$. Following Bozza \cite{bozza2007strong}, we write the result in the following form:
\begin{equation}
    \Delta \tilde{\phi} = \int_{R}^{r_S}b\sqrt{\frac{B(r)}{D(r)\mathcal{R}(r,b)}}dr+\int_{R}^{r_O}b\sqrt{\frac{B(r)}{D(r)\mathcal{R}(r,b)}}dr
    \label{eq:Deltaphi} \, .
\end{equation}
Here 
\begin{equation}
    \mathcal{R}(r,b) = \frac{D(r)}{A(r)}-b^2 \, , \quad
    b = \Big| \dfrac{p_{\tilde{\phi}}}{p_t} \Big|
    \label{eq:Rfunction}
\end{equation}
where $p_t$ and $p_{\tilde{\phi}}$ are the constants of motion associated with the Killing vector fields $\partial _t$ and $\partial _{\tilde{\phi }}$, respectively. For a light ray that comes in from infinity, the constant of motion $b$ is known as the \emph{impact parameter}. The minimum radius $R$ can be expressed by $b$, and vice versa, with the help of the relation
\begin{equation}
    \mathcal{R}(R,b) = 0 \, .
\end{equation}
If $R$ approaches $r_{\mathrm{ph}}$ from above, $\Delta \tilde{\phi}$ goes to infinity, indicating that the light ray makes more and more turns around the center. The corresponding limiting value of $b$ is called the \emph{critical impact parameter} $b_{\mathrm{cr}}$, defined by
\begin{equation} \label{b-cr-definition}
    \mathcal{R}(r_{\mathrm{ph}},b_{\mathrm{cr}}) = 0 \, .
\end{equation}
Our assumptions imply that Eq.~\eqref{eq:Deltaphi} is valid for all light rays with an impact parameter greater than the critical one. Conversely, for light rays with an impact parameter smaller than the critical one, the radius coordinate is monotonic and, thus, there is no minimum value. In this case the right-hand side of Eq.~\eqref{eq:Deltaphi} has to be replaced by one integral from $r_S$ to $r_O$.

A useful parametrization for the impact parameter for this problem is writing it as a perturbation with respect to $b_\textrm{cr}$ in the simple following form:
\begin{equation}
b = b_{\textrm{cr}}(1+\epsilon) \, .
\end{equation}
For $\epsilon$ in the range $[-1,+\infty \, [ \, $, all possible values of the impact parameter are taken into account but, since we are later interested in photons experiencing arbitrarily large deflections and thus passing through a small shell outside the photon sphere, in our treatment the quantity $\epsilon$ is considered small and positive:
\begin{equation}
\epsilon = \frac{b-b_\textrm{cr}}{b_\textrm{cr}} \ll 1 \, .
\end{equation}

The strong deflection limit refers to the case that $R$ approaches $r_{\textrm{ph}}$ and $b$ approaches $b_\textrm{cr}$, namely to the case that the  function $\mathcal{R}$ approaches $0$ and the integrals in Eq.~\eqref{eq:Deltaphi} go to infinity. A simple logarithmic approximation formula for the integrals in $\Delta {\tilde{\phi}}$ has been provided by Bozza \cite{bozza2002gravitational} for the case that both the source and the observer are infinitely distant. This was generalized by Bozza and Scarpetta \cite{bozza2007strong} to the case that $r_S$ and $r_O$ may be finite, which resulted in the equation
\begin{equation} \label{eq:deflection}
     \Delta \tilde{\phi}  = - \tilde{a}\log\frac{\epsilon}{\eta_O\eta_S}+\tilde{a}\log\frac{2\beta_{\textrm{ph}}}{b^2_{\textrm{cr}}}+k_O+k_S \, ,
\end{equation}
where 
\begin{equation}
\eta = 1-\frac{r_{\textrm{ph}}}{r}
\label{eq:eta}
\end{equation}
is a useful dimensionless coordinate.

All other quantities in Eq.~\eqref{eq:deflection} are exclusively associated with the metric coefficients: The radius of the photon sphere is found by solving Eq.~\eqref{eq:Atkinson} and $b_\textrm{cr}$ is defined by Eq.~\eqref{b-cr-definition} which can also be written as
\begin{equation}
    b_{\textrm{cr}} = \sqrt{\frac{D(r_{\textrm{ph}})}{A(r_{\textrm{ph}})}} \, .
\end{equation}
The other quantities in Eq.~\eqref{eq:deflection} are defined by
\begin{equation}    
    \tilde{a} = r_{\textrm{ph}}\sqrt{\frac{B(r_{\textrm{ph}})}{A(r_{\textrm{ph}})\beta_{\textrm{ph}}}} \, ,
\label{eq:tildea}
\end{equation}

\begin{equation}
    \beta_{\textrm{ph}} = \frac{1}{2} r_{\textrm{ph}}^2 \frac{D''(r_{\textrm{ph}})A(r_{\textrm{ph}}) -A''(r_{\textrm{ph}})D(r_{\textrm{ph}})}{A^2(r_{\textrm{ph}})} \, ,
    \label{eq:betaphoton}
\end{equation}
\begin{equation}
    k_i = \int_0^{\eta_i} g(\eta) \, d\eta \, ,
    \label{eq:constantk}
\end{equation}
\begin{gather} \label{eq:g-eta-def}
 g(\eta) = b_{\textrm{cr}} \sqrt{\frac{B(\eta)}{D(\eta)}} \frac{1}{\sqrt{\mathcal{R}(\eta,b_{\textrm{cr}})}}\frac{r_{\textrm{ph}}}{(1-\eta)^2} \, - \\
 - \, \frac{b_{\textrm{cr}}}{\sqrt{\beta_{\textrm{ph}}}} \sqrt{\frac{B(r_{\textrm{ph}})}{D(r_{\textrm{ph}})}} \frac{r_{\textrm{ph}}}{\lvert\eta\rvert} \, , \nonumber
\end{gather}
where we used the index $i$ in order to include both the subscripts $O$ and $S$ denoting the observer and the source, respectively. In order to be consistent with our previous works, in particular with the paper of Tsupko \cite{tsupko2022shape}, we have changed the notation in comparison to the paper by Bozza and Scarpetta \cite{bozza2007strong}. For the reader's convenience, the comparison of the notation is summarized in the Appendix in Table \ref{tab:table}.

Inverting Eq.~\eqref{eq:deflection}, we can express the impact parameter as a function of the azimuthal shift:
\begin{equation}
b = b_{\textrm{cr}} \left\{ 1 + F(r_S)F(r_O) \exp\left(-\frac{\Delta \tilde{\phi}}{\tilde{a}} \right) \right\} \, ,
    \label{eq:impact}
\end{equation}
where
\begin{equation}
    F(r_i) = \sqrt{\frac{2\beta_{\textrm{ph}}}{b^2_{\textrm{cr}}}} \left( 1-\frac{r_{\textrm{ph}}}{r_i} \right) \exp \left(\frac{k_i}{\tilde{a}} \right) \, .
\end{equation}
If we use for each image the \emph{order} $n$, as defined above, and if we denote by $\gamma$ the convex angle between its arrival and departure directions (Fig.~\ref{fig:geometry}), we can express the azimuthal shift as follows:
\begin{equation}
    \Delta \tilde{\phi}  = 
    \begin{cases}
        n\pi + \gamma \qquad  \qquad &\text{for even} \ n \, ,\\        
        (n+1)\pi - \gamma \qquad &\text{for odd} \ n \, .
    \end{cases}
\end{equation}
We remind the reader that the primary or direct image corresponds to $n=0$, while the secondary image corresponds to $n=1$. For values of $n\geq2$, the photons complete at least one full loop around the black hole, and these resulting images are referred to as higher-order (or relativistic) images. In the recent literature about photon rings, the number $n$ is often referred to as ``number of half-orbits'', see, e.g., \cite{johnson2020universal}.

If we consider larger values of $n$, the approximation of the strong deflection limit improves, i.e. the discrepancy with respect to the exact solution becomes smaller and smaller, see \cite{Bozza2010}. The approximation cannot be applied to images of order 0 or 1, but for higher-order images $n \ge 2$, on which we focus in the following, it is usually very accurate.

From now on we assume that the observer is in the northern hemisphere, $0 \le \vartheta _O \le \pi /2$, see again Fig.~\ref{fig:geometry}. As we are free to make a coordinate change $\vartheta \mapsto \pi - \vartheta$, this is no restriction of generality. If we contemplate a narrow emitting ring at radius $r_S$ in the equatorial plane ($\vartheta_S=\pi/2$, $0 \le \phi _S < 2 \pi$) as the light source, the impact parameter given by Eq.~\eqref{eq:impact} becomes a function of the angle $\gamma$, which varies within the range of 0 to $\pi$. More precisely, moving along the ring, $\gamma$ goes from $\pi /2 - \vartheta _O$ to $\pi /2 + \vartheta _O$ and back.

If an observer at radius value $r_O$ sees a certain source under an angle $\theta$ with respect to the radial direction, the conditions of asymptotic flatness Eq.~\eqref{eq:asy} imply that $\theta$ goes to zero as we shift the observer to infinity along a radial line. It is well known and straightforward to verify that then 
\begin{equation}
    b = \underset{r_O \to \infty}{\mathrm{lim}} r_O \, \mathrm{sin} \, \theta = \underset{r_O \to \infty}{\mathrm{lim}} r_O \, \theta \, ,  
\label{eq:btheta}
\end{equation}
This means that for a sufficiently distant observer, the impact parameter $b$ is directly related to an angle in the sky. For this reason, $b$ may be viewed as the radial coordinate on the observer's screen, see Fig.~\ref{fig:geometry}. Moreover, if the observer is at an inclination angle $\vartheta_O$ relatively to the polar axis, and if the limit $r_O \to \infty$ is taken, than the Luminet formula \cite{Luminet1979}, modified in \cite{tsupko2022shape}, provides a means to express $\gamma$ in terms of the polar angle $\varphi$ measured on the observer's screen:
\begin{equation}
\gamma = 
\begin{cases}
     \pi - \arccos{\frac{\sin{\varphi}}{\sqrt{\sin^2{\varphi}+\cot^2{\vartheta_O}}}} \qquad &\text{for even} \ n \, ,\\
     \arccos{\frac{\sin{\varphi}}{\sqrt{\sin^2{\varphi}+\cot^2{\vartheta_O}}}} \qquad  \qquad &\text{for odd} \ n \, ,
\end{cases}
\label{eq:gammavarphi}
\end{equation}
where the $\arccos$ function takes values in the interval $[0, \pi ]$. In the case $\vartheta _O \neq \pi/2$, for each $n$, the angle $\varphi$ runs from 0 to $ 2 \pi$ as we travel around the emission ring. When the angle $\varphi$ goes from 0 to $2\pi$, the angle $\gamma$ takes each value between $\pi /2 - \vartheta _O$ and $\vartheta _O - \pi /2$ twice. Equation~\eqref{eq:gammavarphi} results in a unified expression of the azimuthal shift for both even and odd $n$ that takes the following form \cite{tsupko2022shape}:
\begin{equation}
    \Delta \tilde{\phi}  = (n+1) \pi-\arccos{\frac{\sin{\varphi}}{\sqrt{\sin^2{\varphi}+\cot^2{\vartheta_O}}}} \, .
    \label{eq:azimuthalshift}
\end{equation}
It is worth noticing that for polar observers ($\mathrm{sin} \, \vartheta_O=0$)
this relation simply becomes (see Fig.~1 in Bisnovatyi-Kogan and Tsupko \cite{BK-Tsupko-2022})
\begin{equation}
\Delta \tilde{\phi}  = \left( n + \frac{1}{2} \right) \pi \, .
\end{equation}
In the case of an infinitely distant observer, the function $F(r_O)$ and the constant $k_O$ reduce to the following straightforward expressions:
\begin{equation}
     F(r_O\to +\infty) = \sqrt{\frac{2\beta_{\textrm{ph}}}{b^2_{\textrm{cr}}}} \exp \left( \frac{k_O}{\tilde{a}} \right) \, ,
\end{equation}
\begin{equation}
    k_O = \int_0^1 g(\eta) \, d\eta \, .
\end{equation}
Inserting these expressions and Eq.~\eqref{eq:azimuthalshift} into Eq.~\eqref{eq:impact} gives the impact parameter $b_n$ as a function of the polar angle $\varphi$:
\small\[
b_n(\varphi) = b_\textrm{cr} \left\lbrace1+ \frac{2\beta_{\textrm{ph}}}{b^2_{\textrm{cr}}} \left(1-\frac{r_{\textrm{ph}}}{r_S}\right) \exp\left[\frac{k_O+k_S-(n+1)\pi}{\tilde{a}}\right]\right. 
\]
\begin{equation}
\times \left. \exp\left[\frac{1}{\tilde{a}}\arccos{\frac{\sin{\varphi}}{\sqrt{\sin^2{\varphi}+\cot^2{\vartheta_O}}}}\right]\right\rbrace \, .
    \label{eq:bnformula}
\end{equation}
In the case of a polar observer, $\vartheta _O = 0$, the impact parameter $b_n$ becomes independent of $\varphi$,
\begin{equation}
 b_n = b_\textrm{cr} \left\lbrace1+ \frac{2\beta_{\textrm{ph}}}{b^2_{\textrm{cr}}}\left(1-\frac{r_{\textrm{ph}}}{r_S}\right) \exp\left[\frac{k_O+k_S-(n+\frac{1}{2})\pi}{\tilde{a}}\right]\right\rbrace \, .
\label{eq:bnformulapolar}
\end{equation}
The expression given in Eq.~\eqref{eq:bnformula} describes the shape of the $n$-th image produced by the emission ring ($n \geq 2$). Note that these images are circular only when observed from a polar point of view, according to Eq.~\eqref{eq:bnformulapolar}, and distorted otherwise. Also note that, by Eq.\eqref{eq:bnformula}, $b_n (0)$ and $b_n ( \pi )$ and, thus, the horizontal cross section through the center of the coordinate system ($b = 0$)  of each ring, are independent of $\vartheta _O$; see also Subsection II C of Kocherlakota \textit{et al}. \cite{Kocherlakota-2023-photon-rings}.

We emphasize that for Eq.~\eqref{eq:bnformula} it is of crucial importance that the spacetime is asymptotically flat, according to Eq.~\eqref{eq:asy}, and that the limit $r_O \to \infty$ has been taken. Moreover, we also emphasize that Eq.~\eqref{eq:bnformula} is based on the validity of Eq.~\eqref{eq:Deltaphi}, as it is valid for light rays that start at a radius coordinate $r_S$ ($> r_{\mathrm{ph}}$), go inward to a minimum radius $R$ and then outward to the observer. It is important to realize that in the situation considered here ($r_{\mathrm{ph}} < r_S < r_O$) this is true only for light rays of order $n \ge n_{\mathrm{min}} ( r_S )$, where $n_{\mathrm{min}} (r_S) \to \infty$ for $r_S \to r_{\mathrm{ph}}$. In other words, if the light source is very close to the photon sphere then all light rays up to some very high order $n$ will actually go from the source to the observer without a turning point, namely the radius coordinate will monotonically increase along the entire ray. Therefore, when applying our formalism to light rays of order $n$ we have to make sure that $r_S$ has been chosen such that $n \ge n_{\mathrm{min}}(r_S)$. As $n_{\mathrm{min}} (r_S )$ cannot usually be calculated analytically, we have to check numerically that this criterion is satisfied when applying Eq.~ \eqref{eq:bnformula} to specific examples below.

\section{Examples}
\label{sec:examples}

In this section we specify the metrics we want to use and write explicitly all the relevant quantities. This provides the shape of photon rings \eqref{eq:bnformula} for these metrics.

\subsection{Schwarzschild metric}

The Schwarzschild metric is the unique spherically symmetric solution of Einstein's equation in vacuum. The metric coefficients are
\begin{equation}
A(r) = B^{-1}(r) = 1 - \frac{2m}{r} \, , \qquad D(r)=r^2  \, ,
\end{equation}
where $m$ is the mass parameter with the dimension of a length ($m=GM/c^2$ where $M$ is the mass of the black hole), with the event horizon at the Schwarzschild radius
\begin{equation}
    r_h = 2m \, .
\end{equation}
The radius of the photon sphere is
\begin{equation}
    r_{\textrm{ph}} = 3m 
\end{equation}
while the other quantities of interest (from Eq.~\eqref{eq:tildea} to Eq.~\eqref{eq:betaphoton}) are 
\begin{equation}
    \tilde{a} = 1 \, ,
\end{equation}
\begin{equation}
    b_\textrm{cr} = 3\sqrt{3}m \, ,
\end{equation}
\begin{equation}
    \beta_{\textrm{ph}} = 81m^2 \, .
\end{equation}
The function $g(\eta)$ [see Eq.~\eqref{eq:g-eta-def}] can be integrated analytically giving the constants $k_i$ from Eq.~\eqref{eq:constantk},
\begin{equation}
    k_S = \log{\frac{6}{3-\eta_S+\sqrt{9-6\eta_S}}}
    \label{eq:constantksscw}
\end{equation}
and
\begin{equation}
    k_O = \log{\frac{6}{2+\sqrt{3}}} \, .
    \label{eq:constantkoscw}
\end{equation}
Thus, the curve $b_n(\varphi)$ in its final form of Eq.~\eqref{eq:bnformula} for this specific metric becomes
\begin{equation} \label{eq:schw-photon-ring}
\small
\begin{split}    
   &b_n(\varphi) = 3\sqrt{3}m\left\lbrace1 +\left(1-\frac{3m}{r_S}\right) \left(2+\frac{3m}{r_S}+\sqrt{3+\frac{18m}{r_S}}\right)^{-1}\right. \\
   &\times \left.\frac{216}{2+\sqrt{3}} \exp\left[-(n+1)\pi + \arccos{\frac{\sin{\varphi}}{\sqrt{\sin^2{\varphi}+\cot^2{\vartheta_O}}}}\right] \right\rbrace.
\end{split}
\end{equation}
Equation \eqref{eq:schw-photon-ring} is equivalent to Eq.~(18) of Tsupko \cite{tsupko2022shape}. In particular,
the coefficient in front of the exponential is equivalent to Eq.~(4) in Ref.~\cite{tsupko2022shape}.

\subsection{Reissner-Nordstr\"{o}m metric}

Reissner-Nordstr\"{o}m (RN) metric is a static solution of the Einstein–Maxwell field equations and describes the gravitational field of a spherically symmetric massive object, in particular a nonrotating black hole, endowed with an electric charge, see e.g. \cite{chandrasekhar1998mathematical}.
The metric coefficients are
\begin{equation} \label{eq:RN-metric}
    A(r) = B^{-1}(r) = 1-\frac{2m}{r}+\frac{q^2}{r^2}\ , \qquad D(r) = r^2,
\end{equation}
where $m$ and $q$ ($q^2=GQ^2/4\pi\epsilon_0 c^4$ with $Q$ the electric charge of the black hole) are the mass and the charge parameter with the dimension of a length. 

The horizons are located at
\begin{equation}
    r_h = m\pm\sqrt{m^2-q^2}
\end{equation}
and the outer photon sphere has a radius of
\begin{equation}
    r_{\textrm{ph}} = \frac{1}{2} \left(3m+\sqrt{9m^2-8q^2} \right)\, ,
\end{equation}
see, e.g., \cite{armenti1975existence, dadhich1977timelike, chandrasekhar1998mathematical}.
Again, we list the needed quantities:
\begin{equation}
    \tilde{a} = \frac{3m+\sqrt{9m^2-8q^2}}{\sqrt{2}\sqrt{9m^2-8q^2+3m\sqrt{9m^2-8q^2}}} \, ,
\end{equation}
\begin{equation}
    b_\textrm{cr} = \frac{\left(3m+\sqrt{9m^2-8q^2}\right)^2}{2\sqrt{2}\sqrt{3m^2-2q^2+m\sqrt{9m^2-8q^2}}} \, ,
    \label{eq:bcrRN}
\end{equation}
\begin{equation}
\small
    \beta_{\textrm{ph}} = \frac{\left(3m+\sqrt{9m^2-8q^2}\right)^4\left(9m^2-8q^2+3m\sqrt{9m^2-8q^2}\right)}{8\left(3m^2-2q^2+m\sqrt{9m^2-8q^2}\right)^2} \, .
\end{equation}
The impact parameter of Eq.~\eqref{eq:bcrRN} was first calculated by Zakharov \cite{zakharov1994particle} and can also be found in \cite{zakharov2014constraints, bozza2002gravitational, eiroa2002reissner, perlick2022calculating}.

Unfortunately, this time, the $g(\eta)$ function cannot be integrated analytically and thus the corresponding numerical values are used. On the other hand, if someone is interested in considering only small values of the charge, it is possible to proceed making a series expansion of the integrand $g(\eta)$ in powers of $q$ and end up having the zeroth order $k_i$ being, obviously, Eq.~\eqref{eq:constantksscw} or Eq.~\eqref{eq:constantkoscw} plus analytical corrections of the order $q^2$. This procedure was followed by Bozza in \cite{bozza2002gravitational} in order to provide analytical expressions for $k_S$ and $k_O$. In that paper, these two values were equal (because he considered both source and observer in the asymptotic region) and inserted into his Eq.~(31).

\begin{figure}
    \centering
    \includegraphics[width=0.95\linewidth]{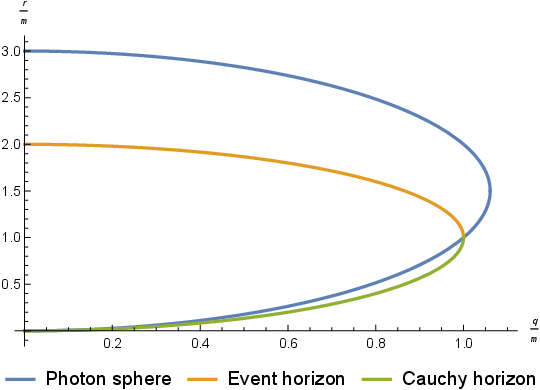}
    \caption{In this graph is represented the behavior of the location of important surfaces in Reissner-Nordstr\"{o}m metric: the Cauchy horizon, the event horizon and the photon sphere as a function of the black hole charge $q$. For $q>m$ there are no horizons. A similar diagram can be found as Fig.~2 in Claudel \textit{et al} \cite{claudel2001geometry}.} 
    \label{fig:ImportantSurfacesRN1}
\end{figure}

As we can see from the previous relations and from Fig.~\ref{fig:ImportantSurfacesRN1}, whereas the photon sphere and the other lensing quantities allow values for the charge $q\leq 3m/2\sqrt{2}$, the horizons exist only for $q\leq m$.

\subsection{Janis-Newman-Winicour metric}

The Janis-Newman-Winicour metric is the unique spherically symmetric and static solution to Einstein's field equation with a minimally coupled massless scalar field as the source. It was found by Janis, Newman and Winicour \cite{janis1968reality} and independently, in a different coordinate representation, by Wyman \cite{wyman1981static}. It was shown by Virbhadra \cite{virbhadra1997janis} that these two solutions are actually the same. Moreover, the same solution was found, again in a different coordinate representation, already in 1948 by Fisher \cite{Fisher1948}.

The metric depends on two parameters, which in the representation now most commonly used are a mass parameter $m$ with the dimension of a length and a dimensionless parameter $\gamma$ (this parameter should not be confused with the angle $\gamma$ from Fig.~\ref{fig:geometry}).
The metric coefficients are
\begin{gather}
    A(r) = B^{-1}(r) = \left(1-\frac{2m}{\gamma r}\right)^\gamma , \\ D(r) = r^2 \left(1-\frac{2m}{\gamma r}\right)^{1-\gamma} \, ,
\end{gather}
and the scalar field is
\begin{equation}
    \Phi(r) = \frac{\sqrt{1-\gamma ^2}}{4 \sqrt{\pi}}\log\left(1-\frac{2m}{\gamma r}\right).
\end{equation}

The parameter $\gamma$ takes values in the range $0 < \gamma \leq 1$. In the limit $\gamma= 1$ the scalar field vanishes and the Schwarzschild solution is recovered. For nontrivial scalar fields the solution does not possess any event horizon, and it describes a naked curvature singularity located at $r=2m/\gamma$ \cite{virbhadra1997nature}.

\begin{figure}
    \centering
    \includegraphics[width=0.95\linewidth]{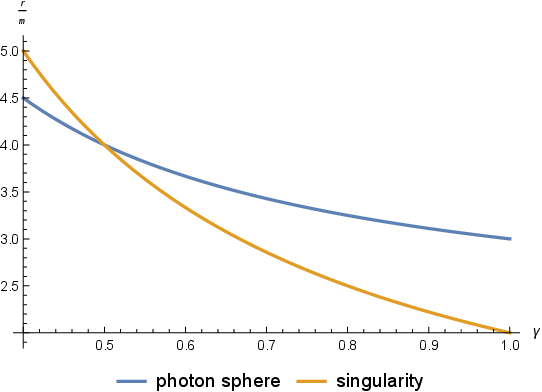}
    \caption{This graph represents the behavior of the photon sphere and the singularity in Janis-Newman-Winicour (JNW) metric as a function of the parameter $\gamma$. For $\gamma>1/2$, although the singularity is always naked, it is below the photon sphere.}
    \label{fig:ImportantSurfacesJNW1}
\end{figure}

The spacetime is asymptotically flat, and for $0.5 < \gamma \leq 1$ it features exactly one photon sphere in the domain $2m/\gamma < r < \infty$ which is located at 
\begin{equation}
r_{\textrm{ph}} = m \left(2+\frac{1}{\gamma}\right) \, ,
\end{equation}
see Virbhadra and Ellis \cite{virbhadra2002gravitational}. Therefore, for these values of $\gamma$ we can apply the formalism developed in Sec.~\ref{sec:shape}. For $0 < \gamma \le 0.5$ there is no photon sphere in the domain $2m/\gamma < r < \infty$ (see Fig.~\ref{fig:ImportantSurfacesJNW1}). In the terminology of Virbhadra and Ellis \cite{virbhadra2002gravitational} the singularity is ``strongly naked'' and the formalism of Sec.~\ref{sec:shape} is not applicable. For this reason, we restrict ourselves to the case $0.5 < \gamma \leq 1$ in the following. As naked singularities are widely assumed to be unphysical, we may join the metric at a radius $r_*$ with $2m/\gamma < r_* < r_{\mathrm{ph}}$ to a regular interior metric that models a star. As long as the surface of this star is not shining, the object would cast a shadow, quite similar to a dark Schwarzschild star with radius between $2m$ and $3m$. The only difference is that here we have a star surrounded by a scalar field, rather than by vacuum.

As regarding the lensing quantities we have
\begin{equation}
    \tilde{a} = 1
\end{equation}
being identical to the Schwarzschild case,
\begin{equation}
    b_\textrm{cr} = \frac{m(2\gamma-1)^{\frac{1}{2}-\gamma}(2\gamma+1)^{\frac{1}{2}+\gamma}}{\gamma} \, ,
\end{equation}
\begin{equation}
    \beta_{\textrm{ph}} = \frac{m^2(2\gamma-1)^{-2\gamma}(2\gamma+1)^{2+2\gamma}}{\gamma^2} \, .
\end{equation}
Also in this case the constants $k_i$ cannot be computed analytically and, as in the Reissner-Nordstr\"{o}m case, we make use of the numerical values,
but again we have the possibility of expanding $g(\eta)$ in powers of $(\gamma-1)$ if we are interested in small deviations with respect to the Schwarzschild metric.

\subsection{Ellis wormhole}

General relativity also admits nontrivial topologies of spacetime, for instance wormholes. The Ellis wormhole \cite{ellis-73} is a particular example of the Morris-Thorne traversable wormhole class \cite{morris1988wormholes} whose metric coefficients are usually written as 
\begin{equation}
    A(r') = B^{-1}(r') = 1 \, , \qquad D(r') = {r'}^2+a^2 \, ,
    \label{eq:ellis}
\end{equation}
where $a$ is the radius of the throat and $r'$ 
runs from $-\infty$ to $+\infty$ spanning the entire spacetime. It is zero at the throat which is a sphere of finite area $4 \pi a^2$, as can be read from the metric. By Eq.~\eqref{eq:Atkinson} this spacetime features exactly one photon sphere which is situated at the throat, $r'_{\mathrm{ph}}=0$. This has the unpleasant consequence that, with $r'$ as the radial coordinate, Eq.~\eqref{eq:eta} does not give an allowed transformation to a new coordinate $\eta$. For this reason, a change of coordinates is necessary if we want to apply the formalism of Sec.~\ref{sec:shape}.

We choose
\begin{equation}
    r = r' + a
\end{equation}
so that the photon sphere is now at $r_{\mathrm{ph}}=a$. The new coordinate $r$ runs over the same range, $\, ] - \infty ,+ \infty [ \,$, as the old one.
With this new radial coordinate the metric becomes
\begin{equation}
     A(r) = B^{-1}(r) = 1 \, , \quad D(r) = r^2 - 2ar + 2a^2 \, ,
\end{equation}
so the conditions of asymptotic flatness Eq.~\eqref{eq:asy} are satisfied.

This particular metric makes the calculations quite easy and provides the following results:
\begin{equation}
    \tilde{a} = 1 \, ,
\end{equation}
\begin{equation}
    b_\textrm{cr} = a \, ,
\end{equation}
\begin{equation}
    \beta_{\textrm{ph}} = a^2 \, ,
\end{equation}
but the major convenience is in the fact that, this time, the function $g(\eta)$ can be integrated analytically giving
\begin{equation}
    k_S = \log \frac{4}{\sqrt{2}+1}\frac{1-2\eta_S+\sqrt{1+(2\eta_S-1)^2}}{\left(1-\sqrt{2}\eta_S+\sqrt{1-2\eta_S+2\eta_S^2}\right)^2}
\end{equation}
and
\begin{equation}
    k_O = \log 2 \, .
\end{equation}
The shape of the $n$th photon ring in polar coordinate Eq.~\eqref{eq:bnformula} assumes, in this metric, the form
\begin{equation}
\footnotesize
\begin{split}
    &b_n(\varphi) = a \left\lbrace 1 + \frac{r_S\left[2a+r_S\left(\sqrt{1+\left(1-\frac{2a}{r_S}\right)^2}-1\right)\right]}{\left[\sqrt{2}a+r_S\left(1-\sqrt{2}+\sqrt{1+\frac{2a^2}{r_S^2}-\frac{2a}{r_S}}\right)\right]^2}\right. \\ 
    \times &\left.\frac{16}{\sqrt{2}+1} \left(1-\frac{a}{r_S}\right)  \exp\left[-(n+1)\pi+\arccos{\frac{\sin{\varphi}}{\sqrt{\sin^2{\varphi}+\cot^2{\vartheta_O}}}}\right] \right\rbrace \, .
    \end{split}
\end{equation}

\section{Constraining the metric by measuring the relative gap between photon rings}
\label{sec:gapparameter}

In this section we discuss the idea of constraining the spherically symmetric metric based on measuring the relative separation between two photon rings, which we will further refer to as gap parameter.

Broderick \textit{et al}. \cite{Broderick-2022-spin} considered photon rings in the case of a Kerr black hole and a polar observer. They proposed using the relative size of a pair of photon rings ($n=0$ with $n=1$, or $n=1$ with $n=2$) to measure the spin of a black hole. Moreover, this issue was also discussed in a paper by Wielgus \cite{wielgus2021photon} for the case of various spherically symmetric metrics. He considered a thin luminous ring of a given radius viewed face-on and calculated the properties of photon rings numerically. In particular, we refer to Fig.~8 in \cite{wielgus2021photon} where the ratio of the $n=2$ and $n=1$ photon ring radii for several metrics as a function of the radius of the emission ring is presented. The ratio of ring radii is also discussed in non-Kerr metrics by Ayzenberg \cite{ayzenberg2022testing}.

Eichhorn \textit{et al}. \cite{Eichhorn-2023-photon-rings} suggested to use the relative separation of photon rings as a probe of regular black holes. These authors discussed the $n=1$ and $n=2$ rings and presented the calculations for a specific emission profile. In particular, we refer the reader to Fig.~3 in \cite{Eichhorn-2023-photon-rings}.

To explain the idea of the gap parameter, we first start from the shadow size. The angular size of the shadow of a spherically symmetric object depends on the parameters of the metric and on the radius coordinate of the observer. For an observer at a large radius coordinate, the angular radius of the shadow is
\begin{equation}
\theta _\textrm{sh} = \frac{b_\textrm{cr}}{r_O} \, .
\end{equation}
We remind the reader that $b_\textrm{cr}$ is the critical value of the impact parameter corresponding to light rays that asymptotically approach the photon sphere, and $r_O$ is the radial coordinate of the observer.

In the case of a Schwarzschild black hole, the only parameter of the metric is $m$, so the impact parameter equals just $m$, which has the dimension of a length, multiplied by a nondimensional constant:
\begin{equation}
b_\textrm{cr} = 3\sqrt{3}m \, .
\end{equation}
Accordingly, the value $\theta _\textrm{sh}$ equals $m/r_O$ multiplied by a nondimensional constant:
\begin{equation}
\theta_\textrm{sh} = \frac{m}{r_O} \, 3\sqrt{3} \, .
\end{equation}

Let us now turn to a more general metric, which depends on a number of parameters $q_0$, $q_1$, $q_2$, .... Without loss of generality, we may assume that all $q_i$ have the dimension of a length. In the case of a spacetime with a nonzero Arnowitt-Deser-Misner (ADM) mass $M$, we may choose $q_0$ to be equal to $m=GM/c^2$. (The ADM mass is well-defined for any asymptotically flat spacetime, see e.g. Misner, Thorne and Wheeler \cite{MTW-1973}, Chapter 19. However, for some spacetimes it is zero. Then we have to choose for $q_0$ some other parameter that occurs in the metric. For example, for the Ellis wormhole, which has vanishing ADM mass, we may use the radius of the throat.) As the impact parameter $b_{cr}$ is a length, it can be written as $q_0$ multiplied by some nondimensional function $\mathcal{F}$ of all metric parameters. Since the function is nondimensional, it can depend only on nondimensional arguments. Therefore, without loss of generality, we can write:
\begin{equation}
b_\textrm{cr} = q_0 \, \mathcal{F} \left( \frac{q_1}{q_0}, \frac{q_2}{q_0}, ... \right) \, .
\end{equation}
Correspondingly, the shadow radius can be written as
\begin{equation}
\theta _\textrm{sh} = \frac{q_0}{r_O} \, \mathcal{F} \left( \frac{q_1}{q_0}, \frac{q_2}{q_0}, ... \right) \, .
\end{equation}
For example, for a Reissner-Nordstr\"{o}m black hole \eqref{eq:RN-metric}, by Eq.~\eqref{eq:bcrRN} we may write
\begin{equation}
\theta _\textrm{sh} =  \frac{m}{r_O} \, \mathcal{F}^{\textrm{RN}}\left( \frac{q}{m} \right) \, .
\end{equation}

However, when observing such a shadow, the only parameter we can discern is its angular radius $\theta_\textrm{sh}$. In the absence of other independent information about the black hole, such as its mass or the radial coordinate of the observer, this radius, in general, does not allow to rule out more complex models of spherically symmetric black holes (beyond the Schwarzschild case). Given a certain radius of the shadow, it can correspond to either a Schwarzschild black hole or to a Reissner-Nordstr\"{o}m one with a different mass or at different distance, or any alternative spacetime which leads to the same size of the shadow. See also the interesting discussion in \cite{Mars-Paganini-2018, Lima-Junior-2021}. For analytical studies of black hole shadows we refer to the recent review \cite{perlick2022calculating}, see also \cite{Cunha-Herdeiro-2018, Vagnozzi-2023-review}.

Now let us consider the lensed image ($n$th photon ring) of the luminous equatorial ring of given radius $r_S$ viewed by a polar observer. For a polar observer, all photon rings will be circles. When limited to the observation of a single photon ring, measuring its angular radius $\theta _n$ carries even more uncertainty than measuring the  shadow boundary. Namely, it will depend again on the parameters of the metric (including the mass in the case that it is nonzero), on the distance to the black hole, and additionally also on the radius $r_S$ of the emission ring:
\begin{equation} \label{eq:single-photon-ring}
\theta _n = \frac{b_n}{r_O} = \frac{q_0}{r_O} \, \mathcal{F}_n\left( \frac{q_1}{q_0}, \frac{q_2}{q_0}, ... \, ; \frac{r_S}{q_0}  \right) \, ,
\end{equation}
compare with Eq.~(1) of Broderick \textit{et al}. \cite{Broderick-2022-spin}.
For example, for a Reissner-Nordstr\"{o}m black hole \eqref{eq:RN-metric}, we have
\begin{equation}
\theta _n =  \frac{m}{r_O} \, \mathcal{F}^\textrm{RN}_n\left( \frac{q}{m}; \frac{r_S}{m}  \right) \, .
\end{equation}
Therefore, the information about the metric will be entangled with the other factors mentioned above. As a result, from the observation of one ring, without additional independent information, it is not possible to rule out alternative metrics beyond the Schwarzschild one.

However, if we would be able to detect a minimum of two photon rings (e.g., $n=1$ and $n=2$, or $n=2$ and $n=3$), then the factor $(q_0/r_O)$ in Eq.~\eqref{eq:single-photon-ring} can be eliminated. For example, if we consider the ratio $\theta_2/\theta_1$ in the case of a spacetime with mass parameter $m \neq 0$, the factor $m/r_O$ will be eliminated \cite{Broderick-2022-spin, wielgus2021photon, ayzenberg2022testing}. 

\begin{figure}
    \centering
    \includegraphics[width=0.47\textwidth]{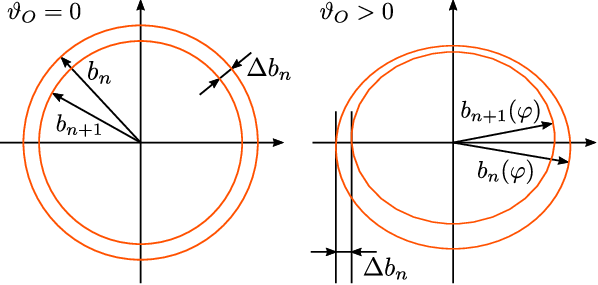}
    \caption{Representation of the appearance of two consecutive photon rings from which the gap parameter is extracted. On the left, it is shown the case of a polar observer. The rings will be perfectly circular and the difference between one ring and another $\Delta b_n$ is constant. The gap parameter is given by $\Delta_n = \Delta b_n/b_n$. On the right, there is the case of an inclined observer with deformed rings and where $b_n$ and $b_{n+1}$ are $\varphi-$dependent. In this case, the points at $\varphi=0$ or $\varphi=\pi$ should be used to measure the gap parameter because the values of $b_n$ in these points of the observer's screen are independent of the observer's viewing angle $\vartheta_\textrm{O}$ [see discussion below Eq.~\eqref{eq:bnformulapolar}]. } 
    \label{fig:gap-parameter}
\end{figure}

Another related possibility is to introduce the gap parameter defined as the relative difference between the radii of two consecutive photon rings normalized by the radius of the bigger one (Fig.~\ref{fig:gap-parameter}):
\begin{equation}
    \Delta_n = \frac{\theta_n - \theta_{n+1}}{\theta_n} 
    = \frac{b_n - b_{n+1}}{b_n}
    = \tilde{\mathcal{F}}_n\left( \frac{q_1}{q_0}, \frac{q_2}{q_0}, ... \, ; \frac{r_S}{q_0}  \right) \, .
    \label{eq:gapparameter-definition}
\end{equation}
For example, for a Reissner-Nordstr\"{o}m black hole \eqref{eq:RN-metric}, we have
\begin{equation}
\Delta_n  = \frac{\theta_n - \theta_{n+1}}{\theta_n}   
= \frac{b_n-b_{n+1}}{b_n} = \tilde{\mathcal{F}}^\textrm{RN}_n \left( \frac{q}{m}; \frac{r_S}{m}  \right)  \, .
\end{equation}

The relative separation $(b_1-b_2)/b_1$ (without using the name gap parameter) was suggested in Eichhorn \textit{et al}. \cite{Eichhorn-2023-photon-rings}.

The gap parameter \eqref{eq:gapparameter-definition} introduced in this way does not depend on the parameter $q_0$ (that is the mass parameter $m$ if it is different from zero) or on the distance to the black hole, which is its main advantage. It depends only on the ratios $q_1/q_0$, $q_2/q_0$ etc. and on $r_S / q_0$, where $r_S$ is the radius of the emission ring. Consequently, if we choose $q_0$ as the length unit and fix all $q_1$, $q_2$, ... , then, by contemplating all possible values of $r_S$, ranging from a certain minimum value to infinity, we can establish a comprehensive range within which the gap parameter can fluctuate for this metric. This, in turn, enables us to differentiate between various spacetimes.

Using the gap parameter allows one to instantly discard, by way of observation, some alternative metrics in comparison with the Schwarzschild spacetime, without knowledge of $m$ and $r_O$, provided that $r_O$ is big enough to make sure that the applied approximations are valid. For each given metric, the allowed interval of the gap parameter can be calculated. Therefore, if the measurement of two photon rings results in a gap parameter outside of this interval, then it unambiguously indicates that the observations are incompatible with this particular metric.\\\\

\section{The gap parameter for higher-order photon rings}
\label{sec:gapparameter-analytical}

In this section, we obtain an analytical expression for the gap parameter for high-order rings using the formulas derived in the previous sections. Since we use the strong deflection limit, the results can only be used for photon rings starting from the $n=2$ ring. For example, for the pair $n=2$ and $n=3$, or for the pair $n=3$ and $n=4$.

\begin{figure*}
    \centering
    \includegraphics[width=0.95\textwidth]{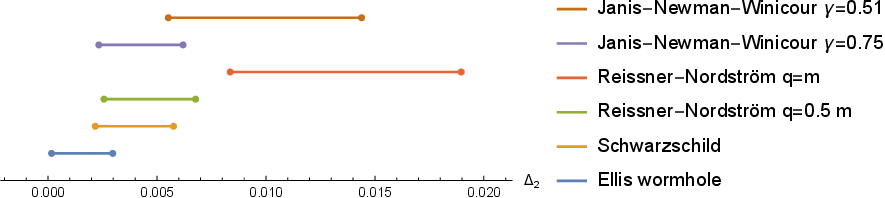}
    \caption{
    Intervals of the gap parameter $\Delta_2$, which is relative separation between $n=2$ and $n=3$ photon rings, for the different spherically symmetric metrics. For each metric, the interval of possible values of the gap parameter is obtained by varying the emissive radius $r_S$ from a chosen minimum value to infinity. For the case of the Schwarzschild, the Reissner-Nordstr\"{o}m and the Janis-Newman-Winicour spacetimes we use the ISCO radius as the minimum value of $r_S$. For the Ellis wormhole, the case is physically different because the source cannot lie at a stable orbit like in the other examples; we use 10/9 as the minimum value for $r_S/a$. The calculations are made using the analytic formula \eqref{eq:gapparameter}. To use the gap parameter, it is sufficient to measure the radii of two subsequent photon rings and calculate the relative separation between them. The resulting number should be compared to this graph. If the number is outside some interval, then the corresponding metric can be discarded.}
    \label{fig:gapfigure}
\end{figure*}

Using our results with the strong deflection approximation, for images of order $n \ge 2$ we can write
\begin{equation}
 \Delta_n   \simeq  f(r_S)\left[1-\exp\left({-\frac{\pi}{\tilde{a}}}\right)\right]\exp\left[{-\left(n+\frac{1}{2}\right)\frac{\pi}{\tilde{a}}}\right] \, ,
    \label{eq:gapparameter}
\end{equation}
where
\begin{equation}
    f(r_S)=\frac{2\beta_{\textrm{ph}}}{b^2_{\textrm{cr}}}\left(1-\frac{r_{\textrm{ph}}}{r_S}\right) \exp\left[\frac{k_O+k_S}{\tilde{a}}\right] \, .
\end{equation}
Coefficients used here and details of their calculation can be found in Secs.~\ref{sec:shape} and \ref{sec:examples}.

In Fig.~\ref{fig:gapfigure} we show the range of variability of the gap parameter for the metrics considered as examples in Sec.~\ref{sec:examples}: the Schwarzschild, the Reissner-Nordstr\"{o}m, the Janis-Newman-Winicour and the Ellis wormhole metrics. If the metric is defined by a single parameter (as in the case of a Schwarzschild black hole or of an Ellis wormhole), then this parameter drops out due to the definition of the gap parameter, see the previous section. If the metric is defined by two parameters (Reissner-Nordstr\"{o}m and Janis-Newman-Winicour metrics), then one drops out and we fix the other in units of the first.

When a metric has been chosen in the above way, we allow values of $r_S/q_0$ between some minimum value and infinity. As indicated above, we choose for $q_0$ the mass parameter $m = GM/c^2$ in all cases where the ADM mass is nonzero. For the Ellis wormhole we choose for $q_0$ the parameter $a$ which determines the radius of the throat. In the case that the ADM mass of the spacetime is nonzero and that there is an ISCO, it is natural to choose $r_{\mathrm{ISCO}}/m$ as the minimum value. For the three metrics where these assumptions are satisfied we have \cite{Kaplan-1949, armenti1975existence, dadhich1977timelike, gyulchev2019image}
\begin{equation}
    r_\mathrm{ISCO} =
    \begin{cases}
    6m &\text{Schwarzschild,} \\
    2m+\rho+\frac{4m^2-3q^2}{\rho} &\text{RN,}\\
    \frac{3\gamma+1+\sqrt{5\gamma^2-1}}{\gamma}m &\text{JNW,}\\
    \end{cases}
\end{equation}
with
\begin{equation}
\small
    \rho=\left(\frac{8m^4-9m^2q^2+2q^4-q^2\sqrt{5m^4-9m^2+4q^4}}{m}\right)^\frac{1}{3} \, .
\end{equation}
In the Ellis wormhole spacetime circular timelike geodesics exist only at the throat (and they are unstable), so there is no ISCO. In this case the choice of the minimum radius is ambiguous. For Fig.~\ref{fig:gapfigure} we have chosen 10/9 as the minimum value for $r_S/a$ so that the starting value of the variable $\eta$ is 1/10.  

In any case, we have to make sure that for all chosen values of $r_S$ a light ray of the relevant order from the source to a distant observer starts with an inward-directed tangent because otherwise our equations are not valid, recall the last paragraph of Sec.~\ref{sec:shape}. In the notation introduced there, our equations can be used for calculating the gap parameter $\Delta _2$ if $2 \geq n_{\mathrm{min}} (r_S)$. This has to be checked numerically. For example, for the Schwarzschild spacetime we find that it is true as long as $r_S \gtrsim 3.018 m$, so by choosing $r_S \ge r_{\mathrm{ISCO}}$ we are on the safe side. Similarly, the criterion is also satisfied for our other examples.

We have to emphasize that the gap parameter $\Delta _2$ is defined for an emission ring of fixed radius coordinate $r_S$, and that its value depends on $r_S$. By contrast, in real observations we expect to see, instead of an emission ring, an extended distribution of emission, an emission \emph{disk}. Strictly speaking, for such a disk there is not a unique gap parameter, because all images get finite angular thickness. However, the images of order $n \ge 2$ of such a disk are so narrow that, for the foreseeable future we do not expect that it will be possible to assign to them, by observation, a finite radial extension. In this sense, the observation will give us a \emph{unique} gap parameter, and if this is outside of the theoretically predicted interval then the corresponding metric can be conclusively ruled out. In the (probably very far) future, it might be possible to resolve the inner structure of higher-order images or, in another words, to assign to higher-order images of accretion disks a finite radial extension. Then one would have to choose the appropriate parts of the neighboring images $n=2$ and $n=3$, corresponding to the same emission radius $r_S$. For example, location of the emission maximum can be used, see Broderick \textit{et al}. \cite{Broderick-2022-spin}.
Then also in this case it would be possible to rule out certain metrics on the basis of observations, even without knowing the details of the distribution of extended emission.

As another word of caution, it should be added that even in the case that an ISCO exists the choice of a minimum value is not unambiguous: If one allows for accretion disks with pressure or viscosity, i.e., for nongeodesic orbits of the emitting matter, then stable circular orbits are possible below the ISCO. Also noncircular, e.g. inspiraling, orbits might be considered. Allowing light sources closer to the photon sphere would result in smaller $\Delta _2$ values. So for the lower bounds in Fig.~\ref{fig:gapfigure} it is important to keep in mind that in the case of the Ellis wormhole it is ambiguous and that for the other examples it is valid as long as we consider emitting matter in stable circular geodesic motion. As, on the other hand, there is no ambiguity as to the maximum radius value, the upper bounds of the gap parameters in Fig.~\ref{fig:gapfigure} are more significant than the lower bounds. If an observed value is above the allowed interval, then it is unambiguously clear that the corresponding metric is excluded by the observation.

From Fig.~\ref{fig:gapfigure} we read, e.g., that for a pressureless and nonviscous accretion disk around an extremal Reissner--Nordstr{\"o}m black hole the possible values of $\Delta _2$ lie entirely above the allowed intervals of four of the other examples. So if one of these values is actually observed, then the other four cases, with arbitrary accretion disks, are definitely excluded. On the other hand, for a Reissner--Nordstr{\"o}m black hole with $q=0.5 m$ the overlap of allowed $\Delta _2$ values with all other chosen examples is quite large, so it is hardly possible to distinguish such a charged black hole from a Schwarzschild black hole by way of measuring the gap parameter.      

We should keep in mind that $\Delta _n$ is a constant only for a polar observer. When the view is not exactly face-on ($\vartheta_O\neq 0$), then the difference between photon rings becomes $\varphi$-dependent. However, since values $b_n(0)$ and $b_n(\pi)$ are the same independently of the observer's viewing angle, the technique described above can still be used if the gap parameter is measured in these points, see Fig.~\ref{fig:gap-parameter}. 
Here it is important to realize that the cross-hairs are not visible in the sky, so the observer needs some additional information to determine which horizontal cross section to consider. (Only the vertical direction is easily determined by the symmetry.) A possible solution to this issue is based on the fact that high-order rings are exponentially close to the boundary of the shadow, which for spherically-symmetric metrics is a circle. Suppose that we observe the rings of order $n$ and $(n+1)$. The $(n+1)$-ring is in fact a compound of the $(n+1)$-ring and all rings of higher order ($n+2$, $n+3$, ...) that we cannot resolve. The $(n+1)$-ring is exponentially closer to a circle than the $n$-ring. Therefore, we can neglect the difference between the $(n+1)$-ring and a circle for the purpose of determining the center of the coordinate system at the observer's screen. In other words, we define this center by treating the $(n+1)$ photon ring as a circle.

Figure~\ref{fig:gapfigure} can be compared with the results of Eichhorn \textit{et al}. \cite{Eichhorn-2023-photon-rings} where they consider regular black holes and find that the modifications due to new physics increase the separation of the photon rings. A similar effect can be seen from Fig.~\ref{fig:gapfigure} where all black-hole solutions are located to the right compared to Schwarzschild. See also \cite{Carballo-Rubio-2023-photon-rings}.\\\\

\section{Conclusions}
\label{sec:conclusions}

The observation and the further analysis of the shadows of the supermassive black holes M87$^*$ and Sgr A$^*$ by the Event Horizon Telescope Collaboration has yielded a lot of information about these compact objects and their accreting environments  \cite{akiyama2019first1, akiyama2019first2, akiyama2019first3, akiyama2019first4, akiyama2019first5, akiyama2019first6, Kocherlakota-2021, EHT-SgrA-2022-01, EHT-SgrA-2022-02, EHT-SgrA-2022-03, EHT-SgrA-2022-04, EHT-SgrA-2022-05, EHT-SgrA-2022-06}. Projects of future observations suggest that within the next decade the detection of the first photon rings may become feasible \cite{johnson2020universal, pesce2021toward, Johnson-2023-Galaxies, Ayzenberg-arxiv-2023}. Higher-order images of the accretion disk may furnish very precise data about black holes and potentially provide new tests of general relativity.

In our study, we have investigated higher-order photon rings around a spherically symmetric compact object. We have considered a circular ring of emission of given radius in the equatorial plane of the compact object and have studied the lensed images of it (Fig.~\ref{fig:geometry}). The use of the strong deflection limit approximation \cite{bozza2007strong, Bozza2010} enables us to derive a fully analytic formula describing the deformed shape of photon rings as seen by an observer at a large distance with any given inclination (Sec.~\ref{sec:shape}). For images of order $n \ge 2$, Eq.~\eqref{eq:bnformula} provides an explicit dependence of the impact parameter $b_n$ on the polar angle $\varphi$ on the observer's sky. The result is written in terms of the metric coefficients and is therefore completely general for all spherically symmetric static spacetimes. As particular examples, we provide the specific form of $b_n(\varphi)$ for the Schwarzschild, Reissner-Nordstr\"{o}m, Janis-Newman-Winicour and Ellis wormhole metrics (Sec.~\ref{sec:examples}).

Further, we have investigated the relative separation between two neighboring photon rings \cite{Eichhorn-2023-photon-rings}, which we refer to as gap parameter, see Eq.~\eqref{eq:gapparameter-definition} for the definition and Fig.~\ref{fig:gap-parameter} for an illustration. We discuss the gap parameter $\Delta_n$ for a general spherically symmetric and static metric depending on several parameters and discuss how it can be used for characterizing the spacetime geometry (Sec.~\ref{sec:gapparameter}). This follows from the fact that the dependence on the mass of the compact object (or, if the latter is zero, on some other characteristic parameter) and on the distance is eliminated. We focus mainly on the gap parameter for higher-order rings where it is possible to proceed fully analytically using the results from Secs. \ref{sec:shape} and \ref{sec:examples}. An analytical formula for the gap parameter $\Delta_n$ with $n \ge 2$ is given by Eq.~\eqref{eq:gapparameter} in Sec.~\ref{sec:gapparameter-analytical}.

Figure~\ref{fig:gapfigure} illustrates the characteristic intervals of $\Delta_2$ (relative gap between $n=2$ and $n=3$ photon rings) for the chosen spherically symmetric and static spacetimes. This visualization demonstrates how a measurement of the gap parameter can, in certain cases, elucidate the spacetime structure or, at the very least, unequivocally rule out some metrics. Namely, when the parameters of the metric are fixed (in units of the mass $m$ if it is nonzero), the gap parameter depends only on the radius $r_S$ of the emission ring. By letting $r_S$ vary between a minimum value (e.g. the radius of the ISCO, if there is any) and a maximal value (e.g. infinity) we obtain the interval of possible values of the gap parameter. If in observations a number outside of this interval is found, then this allows one to unambiguously  exclude the corresponding metric with the chosen parameters from the consideration.

The fundamental prerequisite for measuring the gap parameter is the possibility to detect at least two photon rings which is beyond the current capacity of the Event Horizon Telescope. The next-generation Event Horizon Telescope project and extensions of the very long baseline interferometry technique using space stations hold the potential to provide the necessary sensitivity to achieve this objective \cite{johnson2020universal, pesce2021toward, Johnson-2023-Galaxies, Ayzenberg-arxiv-2023, Gurvits-2022-space, Kurczynski-2022-space}. This advancement could deepen our understanding of spacetime structure and draw attention to potential deviations from general relativity, offering an opportunity to constrain alternative theories of gravity. In addition, photon rings can be used for cosmological studies. It was suggested by Tsupko \textit{et al}. \cite{Tsupko-Fan-BK-2020} to use the shadow of black holes at cosmological distances as a \textit{standard ruler}; see subsequent discussion in \cite{Qi-Zhang-2020, Vagnozzi-2020-concerns, Eubanks-2021, Renzi-Martinelli-2022, Escamilla-Rivera-2022, Cervantes-Cota-2023-review}. This idea can be implemented by measuring the size of the photon rings.

As mentioned before, the idea of constraining the metric by measurements of two or more photon rings has been discussed already previously in the literature \cite{Eichhorn-2023-photon-rings, wielgus2021photon, ayzenberg2022testing, Kocherlakota-2023-photon-rings, da-Silva-2023-photon-rings, Broderick-Salehi-2023-photon-rings, Carballo-Rubio-2023-photon-rings}. We see two advantages in our work. The relative width of the inter-ring gap provides us with a simple visualization of the difference between metrics independent of the common scale of rings given by $(m/r_O)$. Also, we present a fully analytic expression of this parameter for high-order rings based on the strong deflection limit, where the coefficients $k_i$ for some metrics have to be determined numerically.

\begin{acknowledgments}
The authors thank Valerio Bozza for useful comments. The work of F.A. is supported by Università degli Studi di Salerno. F.A. also thanks Eva Hackmann for the kind hospitality at ZARM, University of Bremen. The work of O.Yu.T. is supported by a Humboldt Research Fellowship for experienced researchers from the Alexander von
Humboldt Foundation; O.Yu.T. thanks Claus Lämmerzahl for great hospitality at ZARM, University of Bremen.
\end{acknowledgments}

\appendix

\section{COMPARISON OF NOTATION WITH BOZZA AND SCARPETTA (2007)}

\begin{table}[h]
\caption{\label{tab:table} Comparison of notation in the present paper and in Tsupko \cite{tsupko2022shape} with notation of Bozza and Scarpetta \cite{bozza2007strong} }
\begin{ruledtabular}
\begin{tabular}{p{0.5\columnwidth}>{\centering}p{0.20\columnwidth}>{\centering\arraybackslash}p{0.20\columnwidth}}
Meaning of variable & Notation in this paper and in Tsupko \cite{tsupko2022shape}
& Notation in Bozza and Scarpetta \cite{bozza2007strong} \\ \hline \\

Coordinates & $t, r, \vartheta, \phi$ & $t, r, \vartheta, \phi$  \\\\

Metric coefficients & $A(r)$, $B(r)$, $D(r)$ & $A(r)$, $B(r)$, $C(r)$  \\\\

Angular coordinate in the plane of the ray (in \cite{bozza2007strong} it is equatorial plane of metric) & $\tilde{\phi}$ & $\phi$  \\\\

Radial coordinate of the closest approach & $R$ & $r_0$ \\\\

Radius of the photon sphere & $r_\textrm{ph}$ & $r_m$  \\\\

Impact parameter & $b$ & $u$  \\\\

Critical value of impact parameter & $b_\textrm{cr}$ & $u_m$  \\\\

Radial coordinate of observer & $r_O$ & $D_{OL}$ \\\\

Radial coordinate of source & $r_S$ & $D_{LS}$ \\\\

The quantity in Eq.~\eqref{eq:betaphoton} of this paper& $\beta_\textrm{ph}$ & $\beta_m$, Eq.~(22) \\\\

Function in Eq.~\eqref{eq:Rfunction} of this paper & $\mathcal{R}(r,b)$ & $R(r,u)$, Eq.~(14) \\\\ 

Coefficient of the $\log$ terms in Eq.~\eqref{eq:deflection} of this paper & $\tilde{a}$ & $a$ \\\\

Constant terms in Eq.~\eqref{eq:deflection} of this paper & $k_O \, , \, k_S$ & $b_O \, , \, b_S$ \\
\end{tabular}
\end{ruledtabular}
\end{table}

\end{document}